\documentclass[preprint]{aastex}

\slugcomment{To appear in P.A.S.P.}

\shortauthors{Gizis \& Reid}
\shorttitle{M Subdwarfs}

\begin{document}

\title{HST Observations of M Subdwarfs}

\author{John E. Gizis}
\affil{Five College Astronomy Department, 
Department of Physics and Astronomy, University of Massachusetts,
Amherst, MA 01003\altaffilmark{1}}
\author{I. Neill Reid}
\affil{Department of  Physics and Astronomy, University of Pennsylvania,
209 South 33rd Street, Philadelphia PA 19104-6396}
\altaffiltext{1}{Current Address:  Infrared Processing and Analysis Center, 
100-22, California Institute of Technology, Pasadena, CA 91125 
e-mail:\email{gizis@ipac.caltech.edu}}

\begin{abstract}
\footnotesize
We present the results of an HST snapshot program to search for
very-low-mass stellar companions to nearby M subdwarfs.  None of our 
nine targetted metal-poor primaries have companions more massive than
the hydrogen burning limit, implying that the halo 
binary fraction is equal to or less than the Galactic disk
binary fraction below $0.3 M_\odot$.   In addition, the more distant
tertiary VB12, an sdM3.0 companion to an F subdwarf double, 
is also unresolved.  We show that the relation between WFPC2 F555W 
and F850LP photometry and ground V,I photometry is consistent with 
theoretical expectations.  We also report that two recently observed
Hyades M dwarfs appear single.  
\end{abstract}


\section{Introduction\label{intro}}

The old Population II stellar halo is a fossil relic of 
the formation of the Galaxy.  
Most observational studies to date have concentrated on F and G subdwarfs
near the turnoff of the halo main sequence. The lowest-mass, metal-poor
stars, M subdwarfs, are intrinsically less luminous, but have a higher
local number density. Observations of these stars allow us to probe both
the stellar mass function and the binary formation frequency favoured by
star formation in the early Universe.

There is currently little information on the binary frequency
for the lowest mass stars in the halo.  The formation mechanism(s)
of binaries remains uncertain, with no clear predictions as to 
whether conditions in the young Galaxy would favor 
the production of more or fewer very-low-mass binaries 
relative to the present-day Galactic disk.  
If the binary fraction is high, then the measurements of the
field halo luminosity function \citep{gfb98,gr99} will be in error.
Current transformations of Population II stellar luminosity functions 
into mass functions depend entirely upon theory, since there
are no M subdwarfs with empirical mass measurements.  

In order to search for M subdwarf binary systems, particularly
systems suitable for mass determinations, 
we have obtained Hubble Space Telescope Planetary Camera (HST PC) images
of spectroscopically classified metal-poor M subdwarfs
\citep{g97} with well-determined trigonometric parallaxes.  The targets
were scheduled in Snapshot mode, which allowed only a fraction
of the allocated targets to be observed.  As it turned out, PC images
were obtained of nine targets known to be within 50 parsecs of the Sun.
An additional observation was obtained of VB12 (LHS 541), a more
distant star of special interest since it is the lowest
mass member of a metal-poor triple \citep{vb,sdmsec}.  

We report the results of our search in this paper.  
In Section~\ref{photometry}, we discuss WFPC2 photometry 
for these stars.  In Section~\ref{binaries}, we discuss
the results of our search for binaries.  In Appendix~\ref{hyades},
we report on two Hyades systems observed by HST since the
publication of \citet{hyades2}.

\section{Photometry\label{photometry}}

Our targets were observed using both the F555W and F850LP filters.
Two F850LP images were obtained to allow cosmic ray removal.
The standard HST pipeline processing was used.  
Photometry for the targets was measured on the HST flight system
\citep{hst} including a correction for CTE effects:
$$
mag = -2.5 \log \left({\rm DN}/t_{exp}\right) + {\rm ZP} + 2.5 \log({\rm GF}) 
-2.5 \log \left( 1+ \left(0.04\times Y/800\right) \right)
$$
where in our case ${\rm GF} = 1.987$ for the PC chip, 
${\rm ZP}=21.725$ for F555W, ${\rm ZP}=19.140$ for F850LP,
and the counts are measured in a 0.5\arcsec aperture.  
The uncertainties due to Poisson statistics are less than 0.01 magnitude,
but the uncertainties discussed by \citet{hst} imply there
may be effects that approach $\sim 0.02-0.03$ magnitude.  
The photometry is listed in Table~\ref{table-data},
along with photometry and parallaxes compiled by \citet{g97}.  
Note that the original classification of LHS 407 as sdM5 was based
upon a noisy Palomar 60-in spectrum.  We have obtained a better
spectrum using the Palomar 200-in. and found that the spectrscopic
indices are TiO5$=0.64$, CaH1$=0.60$, CaH2$=0.32$, and CaH3=$0.53$, 
leading to a classification of sdM5.0 but placing it near the (arbitrary)
esdM/sdM border.    

\begin{figure}
\epsscale{0.8}
\plotone{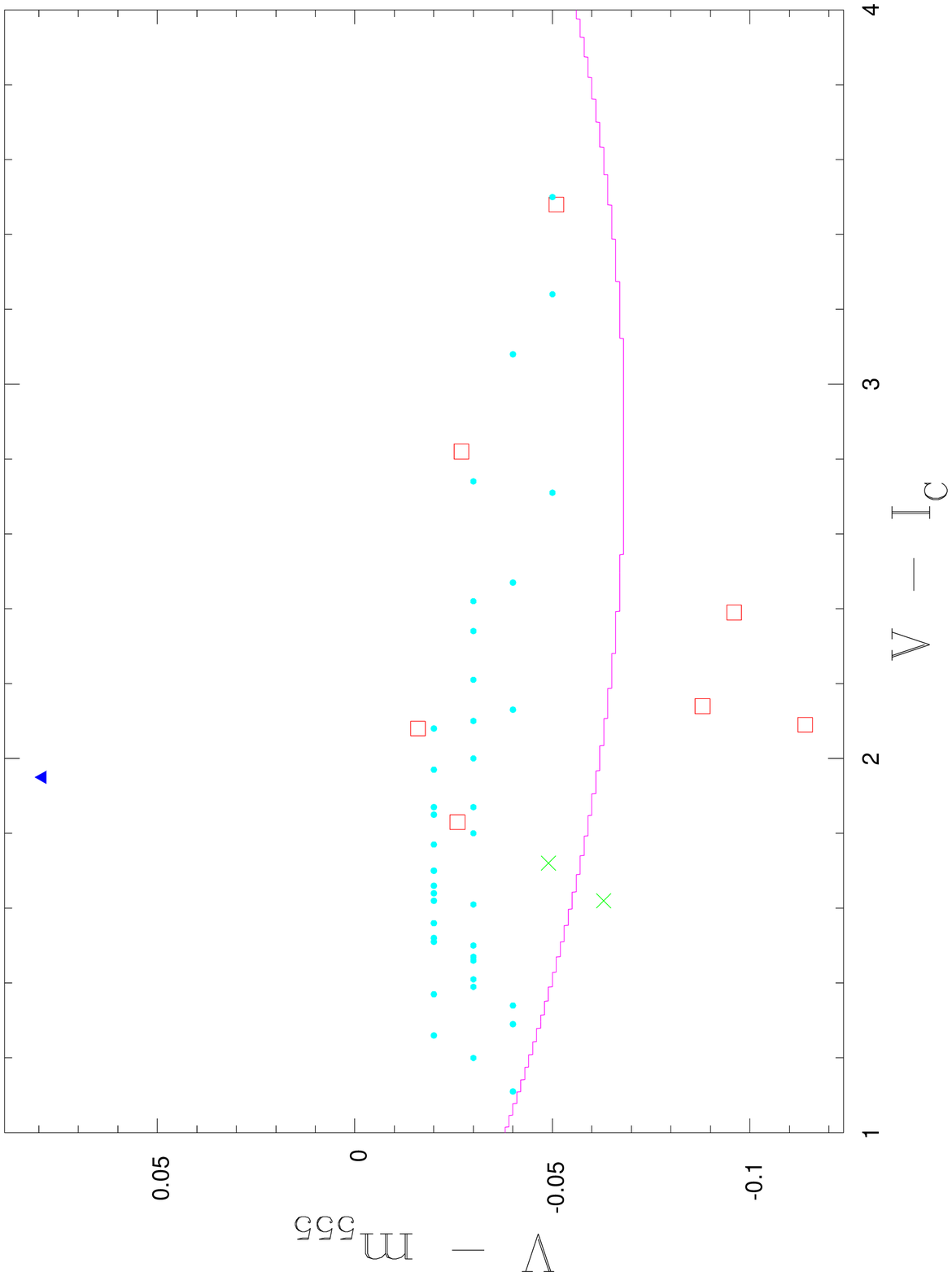}
\caption{Observed F555W and V magnitudes.
Open squares are sdM, the solid triangle is our only esdM (LHS 364), and 
the crosses are late K subdwarfs.   The solid line represents the
synthetic \citet{hst} prediction, while the small points represent
metal poor models of \citet{bcah97}.  The models appear to describe the
data well, but some of the photometry appears suspect. 
\label{fig-vidv}}
\end{figure}

\begin{figure}
\epsscale{0.8}
\plotone{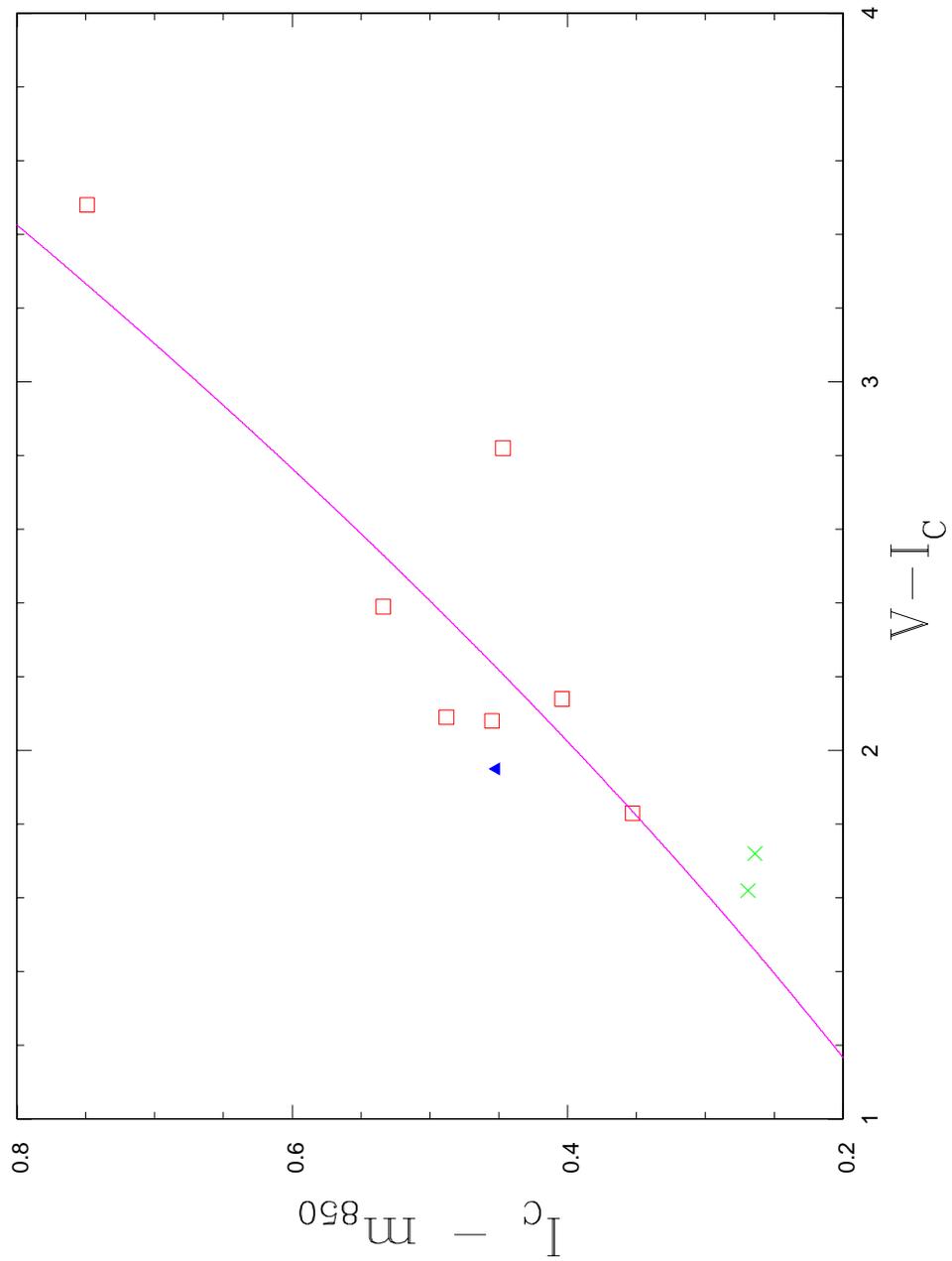}
\caption{Observed F850LP and I$_C$ magnitudes.  The \citet{hst} synthetic 
relation shown as a solid line matches the observations well.  
Symbols are as in Figure~\ref{fig-vidv}
\label{fig-vidi}}
\end{figure}

In Figures~\ref{fig-vidv} and ~\ref{fig-vidi}, we compare the
ground V and I$_C$ photometry compiled by \citet{g97} to our
new HST photometry.   
\citet{l92} estimated that a similar compilation of VI photometry
had uncertainties of 0.05 magnitude.  Also shown in each Figure
are the polynomial fits based on modelling determined by
\citet{hst}.  In the case of the F555W filter, \citet{bcah97} have
calculated both V and F555W magnitudes based on stellar models,
and we also show those calculations in Figure~\ref{fig-vidv}.  
The figures suggest
that the \citet{hst} relation is reliable in transforming F850LP
to I$_C$, while both \citet{bcah97} and \citet{hst} 
are consistent with the F555W-V data.  The observed scatter is consistent
with the probable uncertainties in the ground-based VI photometry.
This result suggests that transformations used to study WFPC2
globular cluster color-magnitude diagrams are reasonable.

\section{Binarity\label{binaries}}

Our observing technique and analysis is essentially identical to 
our previous surveys of Hyades \citep{hyades1,hyades2} and 
field \citep{fieldbin} M dwarfs.  Companions with 
$\delta m_{850}$ of 0, 1, 3 and 5 magnitudes respectively can 
be resolved at  0.09, 0.14, 0.23, and 0.32 arcseconds respectively.  
However, we do not detect {\it any} companions to our target
stars.  

\begin{figure}
\epsscale{0.8}
\plotone{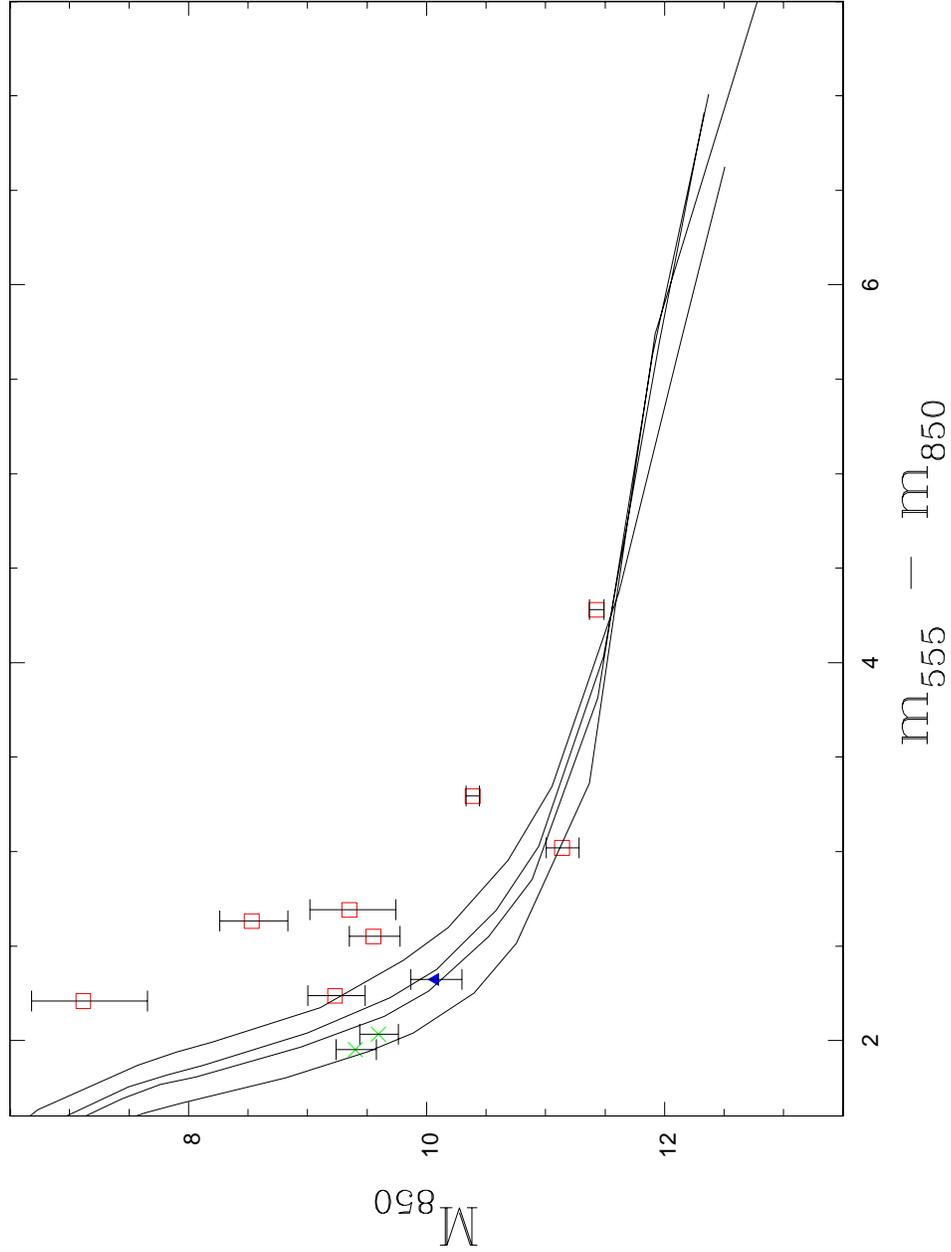}
\caption{Our target stars compared to model calculations.  
Symbols for the targets are as in Figure~\ref{fig-vidv}.  The solid
lines represent \citet{bcah97} models with $[M/H] = -2.0$, -1.5, -1.3,
and -1.0 from left to right.  Since companions with 
$\delta m_{850}$ = 5.0 and 3.0 would be detected at 
0.32 and 0.23 arcsecond respectively.  
The very red tail beyond $m_{555}-m_{850} = 4$ corresponds to masses
less than $0.01 M_\odot$ above the hydrogen burning limit where the
models and colors are very uncertain.  
\label{fig-hstmi}}
\end{figure}

We estimate that our observations are sensitive to stars at the
bottom of the halo main sequence for separations of $>10$ A.U..  
The \citet{bcah97} models predict that end of the metal-poor main sequence
lies at $M_I \approx 14$.  The last 1.7 magnitudes correspond
to masses between 0.083 and 0.090 $M_\odot$, and are predicted
to have very red colors but lie in a regime where the models 
are very uncertain.  These subdwarfs are presumably rare, 
and they have not yet been detected.  Using the \citet{hst} 
transformations, we predict that these dwarfs
have $M_{850} \approx 12.5$, as illustrated in Figure~\ref{fig-hstmi},
but at present there is no empirical verification of the validity of the
color transformations for subdwarfs of such extreme colors.
For Figure~\ref{fig-hstmi}, we have not allowed the I$_C$ to F850LP correction 
to exceed 1.5 magnitudes.
We compare these values to coolest sdM (LHS 377, 
observed $M_{850} = 11.43$) and esdM (LHS 1742, tranformed $M_{850} = 11.1$)
with parallaxes.  \citet{gr97a} and \citet{apmpm2} have found 
extreme M subdwarfs that are slightly cooler than LHS 1742a, but
no parallaxes are yet available.
A multi-epoch HST study of NGC 6397 found no
detected cluster members by $M_I \approx 12.2 - 12.7$ \citep{ngc6397},
which would correspond to $M_{850} \approx 11.5 - 12.0$.  
While the hydrogen burning limit may be at or fainter than this point,
it seems clear that the probability of detecting a subdwarf
in this very small mass range is very low.  Most of our targets
are classified sdM, and therefore are more metal-rich than 
NGC 6397, which may result in a slightly redder, fainter hydrogen
burning limit.  We are sensitive to
$M_{850}=12$ companions as close as 4-10 A.U. for all of the primary targets
except LHS 174 (for which the limit is 15 A.U.), and at distances of
$\gtrsim 12$ A.U. we are typically sensitive to companions as
faint as $M_{850} = 14 -16$.

Since we detect no companions but have only nine nearby targets, the
significance of our result is limited.  
In both the nearby Hyades cluster \citep{hyades1,hyades2} and
field \citep{fieldbin}, we found that 20\% of our HST targets
(which have similar mass but near-solar metallicity) 
were resolved into doubles, corresponding to an overall companion
rate of 35\%.  Thus we expect to observe 1.8 M subdwarf companions
rather than the none actually seen.   
Most of our targets are actually closer than the typical
objects in our previous programs, and we reach the hydrogen
burning limit, so if anything the fraction of observable companions should
be slightly higher.\footnote{We do not expect that the sample is biased
against binaries, which would be overluminous in HR diagrams
and therefore closer to the disk sequence, because the sample
of high velocity parallax stars observed by \citet{g97} include
many spectroscopic non-subdwarfs near the disk main sequence.
It is unclear whether overluminous binaries would be 
preferentially included or excluded by
parallax and spectroscopic studies based on proper motion surveys.}
Our failure to detect any companions suggests
that the binary fraction of M subdwarfs is less than or equal
to that of Galactic disk M dwarfs.  We note also the contrast between
our result and the success found by \citet{keckl}, who found
that three of ten L brown dwarf targets were resolved into 
near-equal-luminosity systems with separations of 5-10 A.U. -- we
would have detected equal-luminosity systems in that range of separations.  

The possibility that the high-velocity, metal-poor stars
have a low binary fraction dates back to \citet{o26}, who
found that high velocity stars were deficient in visual binaries. 
More recently, \citet{aw87} argued that both visual and spectroscopic 
binary fraction of Population II stars is only 40\% that of Population 
I stars.  If so, this is a signature of the formation 
of the Galactic halo --- \citet{stryker} have shown that 
disruption of halo binaries is insignificant.  \citet{stryker} 
and others, however, have argued that the halo binary fraction is
in fact comparable to that of the disk.   
\cite{carney} have found that at least 15\% of their sample
of halo FG subdwarfs are spectroscopic binaries with periods
less than 3000 days.  This is identical to the 14\% spectroscopic
binary fraction of local disk G dwarfs \citep{dm91,mgdm92} in the 
same period range.  Given our small number statistics, our data
are consistent with either scenario.  Our data and the 
G dwarf data taken together strongly suggest that the binary fraction
is not greater than that in the disk.  On the basis of 
imaging of wide binaries in IC 348, 
\citet{dbs99} argue that
loose associations exhibit an excess of binaries with respect
to both denser open clusters (IC 348, Trapezium, the Pleiades) 
and the solar neighborhood.  They suggest that 
the disk binary frequency depends on stellar density within the cluster
(or perhaps some other parameter which also controls the 
density).  If this scenario is correct, and if it applies to the halo, it 
suggests that most halo stars
form in clusters of densities comparable to or greater than the typical
disk star formation region.    

Our results suggest that further study of the binary fraction 
of halo stars of all masses will be profitable.  
A few more M subdwarfs are available for 
study within 50 parsecs; significant improvement in the 
sample size requires extending the sample out to 100 parsecs -- 
increasing numbers of such subdwarfs are being identified.
In addition to imaging, a radial velocity monitoring 
campaign is needed to search for closer 
objects.  Comparison of such data to higher mass halo stars
\citep{carney} and data for disk stars in a variety of enviroments
may provide an important constraint on the formation and subsequent
evolution of the Galactic halo.  The importance of density on
the star formation process and the subsequent evolution of 
halo binary systems also needs to be investigated.  

\section{Summary}

We have found that color transformations of WFPC2 F555W and F850LP to 
ground V,I photometry are consistent with predictions.  
This supports analysis of WFPC2 color-magnitude diagrams of 
globular clusters.  
We find no companions to a sample of nine isolated nearby M subdwarfs
and one tertiary M subdwarf.  A binary fraction as high as that
for similar mass disk M dwarf cannot be ruled out on the basis
of such a small sample, but the lack of observed binaries suggests
that the binary fraction is not high enough to seriously bias
the luminosity function.  
Unfortunately, we have not yet found a system suitable
for mass determinations.

\acknowledgments
This research was supported by 
NASA HST Grant GO-07385.01-96A.

\appendix
\section{Hyades Stars\label{hyades}}

Since our last analysis of our Hyades HST observations \citep{hyades2}, 
two more snapshots have been obtained.  RHy 164 and RHy 326 
have no detected companions.  There are now nine resolved
binaries out of 55, or 16\%; including the three marginally
resolved systems pushes the fraction up to 22\%.  The effect on our
analysis is insignificant.

\begin{table}
\dummytable\label{table-data}
\end{table}


\pagestyle{empty}
\begin{deluxetable}{rrrrrcc}
\tablewidth{0pc}
\tablenum{1}
\tablecaption{Targets}
\tablehead{
\colhead{LHS} &
\colhead{m$_{555}$} &
\colhead{m$_{850}$}&
\colhead{V} &
\colhead{I$_C$} &
\colhead{d (pc)} & 
\colhead{Sp.} 
}
\startdata
169  & 14.179 & 12.146 & 14.13 & 12.41 & 32.4 & esdK7 \\ 
174  & 12.776 & 10.567 & 12.75 & 10.92 & 49.0 & sdM0.5 \\ 
216  & 14.676 & 12.125 & 14.66 & 12.58 & 32.7 & sdM2.0 \\ 
320  & 14.088 & 11.456 & 14.00 & 11.86 & 38.5 & sdM2.0 \\ 
364  & 14.531 & 12.208 & 14.61 & 12.66 & 26.7 & esdM1.5 \\ 
377  & 18.441 & 14.161 & 18.39 & 14.91 & 35.2 & sdM7.0 \\ 
407  & 16.666 & 13.646 & 16.57 & 14.18 & 31.7 & esdM5.0 \\ 
522  & 14.213 & 12.261 & 14.15 & 12.53 & 37.3 & esdK7 \\ 
536  & 14.687 & 12.450 & 14.65 & 99.99 & 44.1 & sdM0.5 \\ 
541  & 16.574 & 13.882 & 16.46 & 14.37 & 80.6 & sdM3.0 \\ 
3409  & 15.187 & 11.893 & 15.16 & 12.34 & 20.0 & sdM4.5 \\ 
\enddata
\end{deluxetable}

\end{document}